\documentclass[authorversion,nonacm]{acmart}

\AtBeginDocument{%
  \providecommand\BibTeX{{%
    \normalfont B\kern-0.5em{\scshape i\kern-0.25em b}\kern-0.8em\TeX}}}

\copyrightyear{2025}
\acmYear{2025}
\setcopyright{cc}
\setcctype{by}
\acmConference[CHI '25]{CHI Conference on Human Factors in Computing Systems}{April 26-May 1, 2025}{Yokohama, Japan}
\acmBooktitle{CHI Conference on Human Factors in Computing Systems (CHI '25), April 26-May 1, 2025, Yokohama, Japan}
\acmDOI{10.1145/3706598.3713520}
\acmISBN{979-8-4007-1394-1/25/04}

\usepackage{makecell}
\usepackage{subcaption}
\usepackage{stfloats} 
\usepackage{graphicx}

\begin{document}

\title[Encountering Friction, Understanding Crises: How Do Digital Natives Make Sense of Crisis Maps?]{Encountering Friction, Understanding Crises:\\How Do Digital Natives Make Sense of Crisis Maps?}



\author{Laura Koesten}
\email{laura.koesten@univie.ac.at}
\orcid{0000-0003-4110-1759}
\affiliation{%
    \institution{Faculty of Computer Science, University of Vienna}
    \city{Vienna}
    \country{Austria}
}

\author{Antonia Saske}
\email{antonia.saske@univie.ac.at}
\orcid{0009-0005-4250-8464}
\affiliation{%
    \institution{Faculty of Computer Science, University of Vienna}
    \city{Vienna}
    \country{Austria}
}

\author{Sandra Starchenko}
\email{sandra.starchenko@yahoo.de}
\orcid{0009-0003-5601-7661}
\affiliation{%
    \institution{Faculty of Computer Science, University of Vienna}
    \city{Vienna}
    \country{Austria}
}

\author{Kathleen Gregory}
\email{k.m.gregory@cwts.leidenuniv.nl}
\orcid{0000-0001-5475-8632}
\affiliation{%
    \institution{Centre for Science and Technology Studies (CWTS), Leiden University}
    \city{Leiden}
    \country{Netherlands}
}

\newcommand\blfootnote[1]{
    \begingroup
    \renewcommand\thefootnote{}\footnote{#1}
    \addtocounter{footnote}{-1}
    \endgroup
}

\settopmatter{printacmref=false}
\renewcommand\footnotetextcopyrightpermission[1]{}


\begin{abstract}
Crisis maps are regarded as crucial tools in crisis communication, as demonstrated during the COVID-19 pandemic and climate change crises. However, there is limited understanding of how public audiences engage with these maps and extract essential information. Our study investigates the sensemaking of young, digitally native viewers as they interact with crisis maps. We integrate frameworks from the learning sciences and human-data interaction to explore sensemaking through two empirical studies: a thematic analysis of online comments from a New York Times series on graph comprehension, and interviews with 18 participants from German-speaking regions. Our analysis categorizes sensemaking activities into established clusters: \textit{inspecting}, \textit{engaging} with content, and \textit{placing}, and introduces \textit{responding personally} to capture the affective dimension. We identify friction points connected to these clusters, including struggles with color concepts, responses to missing context, lack of personal connection, and distrust, offering insights for improving crisis communication to public audiences.
\end{abstract}





\maketitle

\blfootnote{\copyright 2025 Copyright held by the owner/authors. This is the authors' version of the work. It is posted here for your personal use. Not for redistribution. The definitive Version of Record will be published in CHI Conference on Human Factors in Computing Systems (CHI '25), April 26-May 1, 2025, Yokohama, Japan. \url{https://doi.org/10.1145/3706598.3713520}.}

\section{Introduction}

Maps can be vital in crisis communication, as demonstrated during the COVID-19 pandemic, where visualizations were used to inform the public about the scale and impact of the crisis \cite{su2021mental, dong2020interactive, wissel2020interactive}. Designed to convey time-sensitive information \cite{goodchild2010geographic} and guide responses \cite{zhang_mapping_2021}, crisis maps are applicable to many global issues, such as climate change or inflation, where understanding geographical impact and scale is crucial for effective response \cite{griffin_trustworthy_2020}. Although visualizations are considered effective for communicating critical information \cite{field2013you, tait2010presenting, hawley2008impact}, with maps being particularly prominent and memorable \cite{lee_how_2016, magnan2015young, hammond2007text, eden2009patients}, there remains a limited understanding of how public audiences engage with these maps and extract essential information \cite{thorpe2021exposure, franconeri_science_2021}. This understanding can be crucial for informed decision-making in crisis situations \cite{buckingham2009future, kovalchuk2023digitalization}. Existing research tends to focus on experts \cite{chen_rampvis_2020}, leaving a gap in our knowledge about how non-experts, such as younger map viewers, make sense of crisis maps and the challenges they face. This study addresses this gap by examining how young, digitally native viewers \cite{prensky2001digital, Veinberg_digitalNatives_2015} interact with crisis maps, shedding light on their sensemaking processes and identifying friction points, which are inherent to these processes.

\begin{figure*}[ht]
    \centering
    \includegraphics[width=430pt]{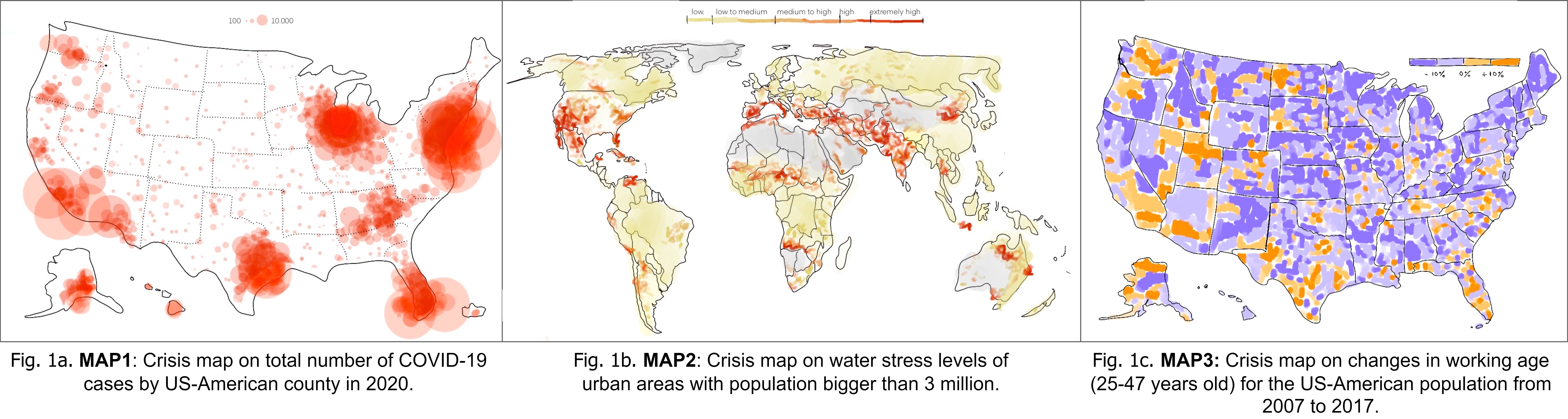}
    \caption{Subfigures 1a-1c depict representations of crisis maps tested in our interview study drawn from a \textit{New York Times} series on graph comprehension. Original map rights belong to the \textit{NYT}. The sketches were created by us for illustration purposes.}
    \label{fig:map-sketches}
    \Description[Subfigures 1a-1c depict representations of crisis maps tested in our interview study drawn from the \textit{New York Times}' graph comprehension series. Original map rights belong to the \textit{New York Times}' Learning Network. These sketches were created by us for illustration purposes.]{Display of three subfigures (1a-1c) depicting representations of crisis maps tested in our interview study. These were drawn from the \textit{New York Times} Learning Network's series on graph comprehension, to which the original map rights belong. We created these sketches for illustration purposes. Crisis MAP1 is a sequential color map with proportional symbols, showing the total number of COVID-19 cases by U.S. county in 2020. Crisis MAP2 is a sequential color map, showing water stress levels in urban areas with populations greater than three million. Crisis MAP3 is a divergent color map, displaying changes in the working-age population (25-47 years old) in the U.S. from 2007 to 2017. Links to the original map sources are listed in Table 2.}
\end{figure*}

We integrate frameworks from the learning sciences \cite{meyer-land2003threshold, goebel2019thresholdconcepts, bloom1956, lee2023affectivelearning} and human-data interaction \cite{koesten_talking_2021} to explore how young public audiences, specifically Digital Natives, engage with crisis maps. We examine their sensemaking in detail through two empirical studies: a thematic analysis of online comments from an educational \textit{New York Times} series on graph comprehension and semi-structured interviews with $18$ Digital Natives from German-speaking regions. In our analysis, emerging sensemaking activities are clustered into \textit{inspecting}, \textit{engaging} with content, and \textit{placing} as also done by \cite{koesten_talking_2021}. We introduce an additional component, \textit{responding personally}, which captures affective activities. A key contribution of our work is the identification of friction points within the sensemaking process, which includes struggling with color encoding, missing context, lacking connection, and distrusting the crisis map. We discuss the implications of these friction points for the design and usage of crisis maps, offering insights that are essential for improving crisis communication with public audiences.

\section{Background}
Crisis maps serve as crucial tools for conveying information about the nature and scale of crises \cite{zhang_mapping_2021, grandi2021geovis}, thereby aiding informed decision-making \cite{du_crisis_2021, dong2020interactive}. While they are vital for understanding crises, their impact on viewers can vary significantly \cite{fan_understanding_2022, kostelnick_cartographic_2013}. Despite the importance of clarity and accuracy in these maps \cite{lee_how_2016}, existing studies on crisis visualization often focus on experts \cite{chen_rampvis_2020}, while neglecting public audiences, such as young, digitally native viewers who are key actors in sharing and consuming online information \cite{lusk2010digital}. We draw on the learning sciences and work in human-data interaction to frame sensemaking as a process that can be deconstructed into sensemaking activity clusters \cite{koesten_talking_2021}, that encompasses both cognitive and affective activities \cite{lee2023affectivelearning, schwartmann2010thresholdconcepts}, and during which friction may occur \cite{goebel2019thresholdconcepts}.

\subsection{Maps in Times of Crises}
\label{crisis-maps}
There are many current issues that may be called crises, including climate change, inflation, and health emergencies. These problems have global effects, and addressing them requires an understanding of the geographical impact and the scope of the situation \cite{griffin_trustworthy_2020}. The media can be central to dealing with and overcoming crises, for instance by enabling effective crisis communication \cite{su2021mental, ferrara2020misinformation} or facilitating discussions about measures to counter and respond to the causes of crises \cite{PACE-Resolution_mediaCrises_2022}. One communication tool in the media are crisis visualizations, which are visual representations of data in potentially threatening circumstances \cite{zhang_mapping_2021}. They provide insights into the nature and scale of a crisis, help identify trends and patterns, and support informed decision-making \cite{du_crisis_2021}. Crisis visualizations address different issues, such as epidemics \cite{fan_understanding_2022, cay_understanding_2020, juergens_trustworthy_2020}, natural disasters \cite{thompson_influence_2015, padilla2020powerful}, and social issues \cite{dobovs2023visualizing, ballas2018analysing}, offering real-time information and engaging the public \cite{middleton2013enhancing, sood1987news, powell2015clearer, bell2011media, kampf2013transforming, singer2018likewar}.

Crisis maps, which are one type of crisis visualization, use geo-referencing to display crisis-related data for communication purposes \cite{grandi2021geovis}. In crises, they can be provided by media outlets, institutions, or private map designers \cite{hullman_benefitting_2011, fang_evaluating_2022, mayr_trust_2019}. Unlike general maps, which may serve broader purposes such as navigation or education, crisis maps are defined through their time-critical data \cite{goodchild2010geographic} and their purpose-driven design which aims to inform and guide crisis responses \cite{zhang_mapping_2021}. Crisis maps were for example central during the COVID-19 pandemic, as they displayed case numbers and fatalities across regions \cite{ferrara2020misinformation, su2021mental, dong2020interactive} to inform the public about the pandemic's scale and impact, proving essential for understanding its global reach \cite{wissel2020interactive}. 

It is thought that crisis maps should be accurate, comprehensive, and clear to facilitate viewer understanding \cite{lee_how_2016}, yet their comprehensibility has been criticized several times \cite{fan_understanding_2022, du_crisis_2021, thompson_influence_2015, fang_evaluating_2022}. In crisis communication, the complexity and dynamism of crisis data pose design and perception challenges \cite{maher2020covid19, pickles2021covid}. Factors like cartographic design choices \cite{macdonald-ross_how_1977, tversky2000some, winn1987communication} and the viewer's personal stance, which can influence trust and emotional reactions \cite{davis2020pandemics, garfin2020novel, davis2014we}, affect map understanding. Investigating the perception of crisis maps is therefore crucial, given that making sense of these maps can play a critical role in times of crisis \cite{ohme_attenuating_2021}.

\subsection{Investigating the Perspective of Young and Digitally Native Crisis Map Viewers}
Crisis visualization viewer groups range from lay viewers \cite{lisnic_misleading_2022, zhang_mapping_2021} to experts \cite{thompson_influence_2015, kostelnick_cartographic_2013}, each with different needs. Thus, viewer characteristics and needs should be considered, both when designing crisis maps and also when investigating their perception \cite{kostelnick_cartographic_2013, zhang_mapping_2021}. The retrieval of information from crisis maps made for experts demands higher visual data literacy \cite{boy2014principled}, that cannot necessarily be expected from lay viewers. Previous studies often focused on experts with advanced skills, and for example, evaluated tools and simulation possibilities \cite{kostelnick_cartographic_2013, chen_rampvis_2020}, or risk perception in crisis visualizations \cite{thompson_influence_2015}. 

In our investigation of crisis map sensemaking, participants belonged to a specific viewer group, which can be defined through their age and digital skills. Our participants were young, digitally native crisis map viewers. Digital Natives are a particularly interesting viewer group to investigate, as they are influential sharers and consumers of online information \cite{lusk2010digital, autry2011digital}. By definition, \textit{Digital Natives} were born after 1980 and, though they are not a monolithic group \cite{correa2016digital}, tend to have a preference for using new technology \cite{helsper2010digital, prensky2001digital}. Existing work shows that they process information quickly, often prefer graphics over text \cite{prensky2001digital} and tend to have high digital skills due to their generational context \cite{underwood2007rethinking}. Digital Natives tend to rely heavily on online sources \cite{Veinberg_digitalNatives_2015}, which is why their efficient map comprehension is crucial for informed decision-making in (future) crises \cite{lusk2010digital, buckingham2009future, kovalchuk2023digitalization}. It is crucial to educate viewers to be critical consumers of visual media \cite{muehlenhaus2014going}, as the online dissemination of misinformation and conspiracy theories \cite{orso2020infodemic, meese2020covid19} can significantly influence behaviors and attitudes during crises \cite{zhang_mapping_2021, zhang_visualization_2022}. Online, such as on social media platforms, crisis maps may inadvertently contribute to the spread of misinformation \cite{lee_viral_2021, lisnic_misleading_2022, lupton_learning_2021}, as has been shown in discussions about climate change and COVID-19 \cite{lupton2013moral, klemm2016swine}.

\subsection{Framing Sensemaking as a Learning Process consisting of Activity Patterns}
From a human-computer interaction perspective, sensemaking involves the process of constructing meaning from information by assembling pieces into a coherent concept \cite{russell1993cost, blandford2010interacting}, encompassing both cognitive and social dimensions \cite{russell2008sensemaking}. According to the data-frame theory \cite{klein2007dataframe}, sensemaking is described as an iterative process influenced by contextual elements such as past experiences, individual perspectives, and prior knowledge, highlighting that it is inherently shaped by these factors. In relation to crises, sensemaking has been similarly described as a continuous effort to interpret and assign meaning to information \cite{wozniak2016ramparts, goyal2013effectsandsensemaking}, with personal responses playing a significant role \cite{zhou_spotlight_2023}.

When viewers engage with crisis maps -- or other information sources -- the sensemaking process is complex \cite{thompson_influence_2015}, context-dependent \cite{cleveland1993}, and iterative \cite{russell2003learning}, with visualizations shown to be able to support sensemaking efforts \cite{goyal2013effectsandsensemaking}. While visual exploration in sensemaking has been well-researched \cite{kang2012sensemakingtasks, yalcin2018visualexploration}, there is a lack of in-depth exploration of its dimensions when it comes to crisis maps, as prior research has primarily focused on map design. Previous studies have assessed how specific map properties affect viewer's risk perception, comprehension, and preferences \cite{bostrom2008visualizing, cao_is_2016, thompson_influence_2015, fang_evaluating_2022, zhang_mapping_2021}. For example, it was shown that combinations of color tones with map types can influence risk communication in maps that display data on COVID-19 \cite{fang_evaluating_2022}, or that different cartographic risk representations influence viewers' decision-making \cite{cheong2016evaluating}, and guidance on effective risk map design has been offered accordingly \cite{maceachren2012visual, monmonier_how_2018, du_crisis_2021, zhang_mapping_2021, xiong_examining_2019, fan_understanding_2022}.

We integrate research in human-data interaction \cite{koesten_talking_2021} with theories from the learning sciences \cite{meyer-land2003threshold, goebel2019thresholdconcepts, bloom1956, lee2023affectivelearning} to create a framework for the sensemaking of crisis maps. By incorporating the learning sciences, we gain a nuanced perspective on the viewer's experience during sensemaking. This approach emphasizes the processual nature of seeking and acquiring information, recognizing challenges -- which previous research has questioned, particularly regarding their influence on efficient comprehension in sensemaking \cite{koesten_talking_2021, boukhanovsky2017uncertainty} -- as integral components of the overall process. This integration also introduces established terminology for different viewer activities and their cognitive and affective dimensions that come into play during sensemaking. We expand on this by incorporating a data-centric sensemaking framework \cite{koesten_talking_2021} that outlines specific data-related sensemaking activities. Based on this framework, we describe sensemaking as a complex and, at moments, frictional process that consists of cognitive and affective activities, which can be grouped into sensemaking activity clusters. 

\subsubsection{Deconstructing Sensemaking into Activity Patterns} Sensemaking is considered an iterative process of assembling pieces into understanding, involving different dimensions \cite{goyal2013effectsandsensemaking, russell2008sensemaking, klein2007dataframe}. Sensemaking has been explored as a collective process in crisis-related scenarios, such as when fragmented information is discussed online \cite{zhou_spotlight_2023, dailey2015collectivesensemaking}. It has been shown to involve categories such as understanding the causes, impacts, and solutions of crises, as well as personal responses to them \cite{zhou_spotlight_2023}. While our approach to deconstructing sensemaking addresses these categories, it provides a more granular structure by describing sensemaking across activities, patterns, and clusters. 

In \cite{koesten_talking_2021}'s framework for data-centric sensemaking, common patterns of cognitive and physical actions are outlined. Drawing on a mixed-methods study of interviews and screen recordings, they identified three sensemaking activity clusters, each with specific activity patterns and data attributes: \textit{Inspecting}, where viewers gain an overview of the data by considering attributes like topic, title, and structure; \textit{Engaging with content}, involving simple analysis and questioning uncertain data elements; and \textit{Placing}, where viewers relate the data to different contexts. The sensemaking activity patterns emerged from studying other data-centric work practices \cite{koesten_talking_2021} but are one of few studies focusing on data specifically as opposed to other or mixed information sources. In our study, the framework is applied and compared to the sensemaking of crisis maps. Using this framework, the different rhythms of people's sensemaking processes \cite{wozniak2016ramparts} -- indicating that sensemaking does not follow a uniform pace -- can be described by grouping sensemaking activities into patterns, and patterns into clusters. 

\subsubsection{Sensemaking as a Learning Process featuring Affective Activities} 
Communicative visualization design has already been approached as a learning design problem, where the visualization viewer is equated with the student and the designer with the teacher \cite{lee2022learningobjectives}. A well-established framework from the learning sciences, also recognized in visualization research for its ability to support a differentiated approach to viewer engagement and needs, is \textit{Bloom's Taxonomy of Educational Objectives for Knowledge-Based Goals} \cite{bloom1956}. This taxonomy encompasses cognitive, affective and psychomotor domains, and breaks learning down into activities, which are most commonly used for cognitive intents \cite{lee2023affectivelearning}. Similarly, Wiggins and McTighe \cite{wiggins2005} deconstruct the process of understanding into six facets, including ability categories closely tied to personal responses: empathy, perspective and self-knowledge. Recognizing the complexity of learning as a multifaceted process, subsequent work has emphasized its iterative and simultaneous nature. Central to our approach is the learning sciences theory of \textit{Threshold Concepts}, which describes learning as an iterative interplay involving simultaneous cognitive, affective, and social activities \cite{schwartmann2010thresholdconcepts}. We incorporate the theory of Threshold Concepts, as it is particularly suited to describe learning complex topics, which are, for example, transformative and integrative \cite{meyer-land2003threshold}, like crisis issues. 

The inclusion of affective activities alongside cognitive ones was recognized in Bloom's original taxonomy \cite{bloom1956} and further expanded upon in its revised version \cite{anderson2001bloomrevised}. This perspective aligns with recent data visualization research, which emphasizes that viewer responses to visualizations often extend beyond purely cognitive domains \cite{lee2022learningobjectives}. Affective factors are often stigmatized and hard to measure because they focus on moods, attitudes, or feelings and develop over undefined periods \cite{lee2023affectivelearning}. In 2023, Bloom's Taxonomy was adapted by Lee and colleagues to address affective visualization intents by thematically analyzing interview codes \cite{lee2023affectivelearning}. Here, to describe affective sensemaking activities that emerged in personal responses, we use the terms proposed by Lee et al. for \textit{Bloom's Affective Taxonomy}, which are \textit{perceive}, \textit{respond}, \textit{value}, \textit{believe} and \textit{behave}.


\subsubsection{Describing Friction Points in Sensemaking}
Alongside cognitive and affective activities, we investigate friction points that arise during crisis map sensemaking, questioning what these points reveal about the sensemaking process \cite{koesten_talking_2021, boukhanovsky2017uncertainty} when recognized and examined as inherent to it. To contextualize friction points as part of the sensemaking process, we draw on the theory of Threshold Concepts, which describes a transformative phase in learning that bridges existing and new knowledge \cite{goebel2019thresholdconcepts}. Threshold Concepts deal with challenging or counter-intuitive knowledge \cite{meyer-land2003threshold} and do not expect learners to always leave a learning process successfully by having fully acquired a concept. Instead, there is a transformative stage, the liminal space, where learners may also get ``stuck'' \cite{goebel2019thresholdconcepts}. A review of 60 papers which apply Threshold Concepts in various contexts highlights the theory's strength in identifying specific troublesome points in learning \cite{correia2024pitfalls-threshold}. Instead of using the term ``troublesome'', as typical in the theory of Threshold Concepts, we use ``friction points'' to emphasize the dynamic nature of sensemaking and avoid negative connotations. To describe friction in crisis map sensemaking, we view it as part of processual learning, which may be placed in the liminal space -- a transitional and often uncomfortable phase where individuals grapple with concepts, revisit ideas, and may feel uncertain or doubt their ability to progress.

\section{Methodology}
Two studies were conducted to examine the sensemaking of young, digitally native viewers. Thematic analysis was first applied to online comments on $13$ crisis maps from an educational series on graph comprehension. This series, part of the \textit{New York Times} Learning Network, was selected due to its six-year history promoting data literacy and critical thinking among young audiences. Informed by insights from analyzing comments in this series, $18$ interviews with a more in-depth scope, which featured three of the prior crisis maps, were conducted and analyzed. This mixed-methods approach generated complementary datasets, offering different insights into how two viewer groups of Digital Natives make sense of crisis maps and what challenges they encounter. Both studies were analyzed using a primarily inductive thematic analysis, complemented by deductive elements drawn from existing literature. The coding process (see Section \ref{meth-analysis}) involved iterative refinement of code names to ensure alignment across datasets, without directly comparing them. This approach also enabled a targeted synthesis of friction points in sensemaking.

\subsection{Thematic Analysis of Online Comments}
For the investigation of crisis map sensemaking by young, digitally native viewers, we conducted a thematic analysis of comments from the \textit{New York Times}' Learning Network series called "What's Going on in this Graph?"\footnote{Series website: \url{https://www.nytimes.com/column/whats-going-on-in-this-graph}.}. This series is publicly accessible, but explicitly aimed towards U.S. students in high school contexts. Each week a graph is provided for debate among registered users, who can answer structured questions on the graph. One week after the posting of a graph, there is a ``reveal session'', where experts provide an analysis and interpretations of the graph. The four questions\footnote{The implementation of the series' questions is exemplified in the graphs linked in Table S1 of the supplementary material, which includes all 13 graphs used in the thematic comment analysis.} posed for each graph are as follows:

\begin{enumerate}
    \item What do you notice? 
    \item What do you wonder?
    \item How does this map relate to you and the society you live in?
    \item What's going on in this graph? Create a catchy headline that captures the graph’s main idea.
\end{enumerate}

Participants in the comment section of the series are primarily U.S. high school students. Comments, sourced from the public series, could be submitted individually or in classroom settings. While some commenters included personal details, this was optional and inconsistently done. Given this, direct authorship and location of the comments cannot be verified. To safeguard anonymity, identifying details were excluded during data processing. Comments were treated with respect for contributors, and identifying information was removed to maintain privacy. In line with ethical considerations for online research \cite{kozinets2010ethics, hookway2008consent}, the analysis of publicly available comments did not necessitate formal ethical approval. 

We analyzed a sample of $13$ crisis maps, drawn from the series, which align with our definition of crisis maps as outlined in Section \ref{crisis-maps}. From over $119$ visualizations published at the time of our comment analysis, $27$ maps were identified, out of which we chose $13$ maps, detailed in Table S1 of the supplementary material. These $13$ maps represent four key map types, according to the classification proposed by Munzner \cite{munzner_visualization_2014}: divergent color maps, sequential color maps, categorical color maps, and proportional symbol maps \cite{zhang_mapping_2021, munzner_visualization_2014}. For the $13$ graphs relevant to our research, we retrieved the series' online comments in December 2022 using the \textit{Selenium} crawler in a custom Python script. To ensure semantic quality, comments were required to be at least $100$ words, guided by the \textit{NYT} series' four-question prompt designed to encourage thoughtful engagement. The word limit was informed by a manual review of comment lengths to balance the inclusion of high-quality, meaningful responses with slightly shorter ones that might reveal frictional sensemaking. From the $13$ graphs studied, we selected MAP4, the graph with a median number of comments (n=$763$), to determine the length of the comments included in our analysis. The review showed that most thoughtful comments were around $112$ words or more, leading to the slightly lower limit of $100$ words. Further, the comments had to be posted prior to the series' ``reveal session''. For the thematic analysis of the comments, we used \textit{Atlas.ti} to code open-ended text-based data (see Section \ref{meth-analysis}).

\begin{table*}[b]
\centering
\caption{Overview of characteristics for 18 interview participants (ages 18 to 28)}
\begin{tabular}{l l c l l}

\makecell[hl]{\textbf{Age} \\ \textbf{range}} &  \makecell[hl]{\textbf{Country of} \\ \textbf{residence}} & \makecell [hl]{\textbf{\# of}\\\textbf{particip.}} & \makecell[hl]{\textbf{Represented highest} \\ \textbf{education backgrounds (\#)}} & \makecell[hl]{\textbf{Represented sectors} \\ \textbf{of occupation (\#)}} \\ \hline
18-20 & Germany & 3 &  \makecell[hl]{Certificate of secondary education (2),\\high school education (1)} & \makecell[hl]{Cultural/creative (1), high school\\ student (1), unemployed (1)} \\ 
18-20 & Austria & 1 & Apprenticeship & Health \\ \hline
21-23 & Germany & 2 & Apprenticeship (2) & Civil (1), craft/industrial (1) \\ 
21-23 & Austria & 3 & \makecell[hl]{Bachelor (1), certificate of secondary\\education (1), high school graduation (1)} & \makecell[hl]{Academical (1), civil (1), cultural/\\creative (1)} \\ 
21-23 & France & 1 & High school graduation & Cultural/creative \\ \hline
24-26 & Germany & 5 & \makecell[hl]{Apprenticeship (1), high school\\ graduation (3), master (1)} & \makecell[hl]{Academical (1), craft/industrial (3),\\
cultural/creative (1)} \\ 
24-26 & Austria & 1 & Apprenticeship & Craft/industrial \\ \hline
25-28 & Germany & 1 & Master & Cultural/creative \\ 
25-28 & Austria & 1 & Master & Health \\ 
\end{tabular}
\label{tab:interview-participants}
\end{table*}

\subsection{Interview Study}

The semi-structured interviews built on the thematic comment analysis, delving deeper into sensemaking and its frictions. We chose three crisis maps for in-depth exploration, that had also been featured in the \textit{NYT} series on graph comprehension and were included in our comment analysis. These crisis maps were chosen based on the assumption that the editors of the \textit{NYT's Learning Network} deemed them relevant and suitable for younger audiences. We interviewed $18$ participants, aged between $18$ and $28$, who were proficient in either English or German. As the \textit{NYT's} graph comprehension series targets high school students but is also applicable in college contexts \cite{nyt2024learningnetwork}, selecting participants within this young audience range was considered appropriate. 

Interview participants were recruited through an open call on social media platforms, specifically \textit{Instagram} and \textit{Facebook}, and supplemented by snowball sampling to ensure they met the target demographic while representing different educational backgrounds and different professional fields. Information on the distribution of regions of residency, professional backgrounds and fields of occupation or study can be found below in \autoref{tab:interview-participants}. As the graphs in the \textit{NYT} series were presented in English, we provided translations of the textual elements for non-native English speakers. Participants were encouraged to engage with the translated versions if they felt more comfortable using them. Participants gave consent to participate in the interview study; they were not remunerated for their participation. The interviews were conducted online via \textit{Zoom} in May and June 2023, lasting between $30$ and $60$ minutes, with a median duration of $49.5$ minutes. Ethical approval for the study was granted by the University of Vienna's ethics committee under reference number $00937$.

We discussed three crisis maps with each participant. Of the $13$ crisis maps (see Table S1 in the supplementary material) analyzed in the comment analysis, we selected three to present to each interview participant. These maps were chosen by first identifying three distinct and prevalent crisis topics (public health, climate change, and economic crisis) and then selecting three different map types (a proportional symbol map, a divergent choropleth map, and a sequential choropleth map). From this pool, we chose the most commented-on maps that matched these criteria. Due to copyright restrictions, we cannot include the original crisis maps in this article, but we provide abstracted versions of the visualizations which we used in the interviews (see \autoref{fig:map-sketches}).

The interviews were structured around a think-aloud task and included questions on sensemaking informed by the prior comment analysis. First, participants were introduced to the study topic, asked for their consent, and requested to share demographic information. Next, they were shown three crisis maps and encouraged to perceive and interpret each one. In follow-up questions, interviewees were asked about their interpretations of the maps, their familiarity with the depicted crisis issues, and whether they found the maps helpful in conveying risk. Finally, they were invited to provide critical feedback on the comprehensibility of the crisis maps, including their overall design and effectiveness in conveying information. The interview schedule is attached in the supplementary material (see Table S2). Interviews were transcribed and analyzed using the qualitative text data analysis software \textit{Atlas.ti}, following the procedure described in Section \ref{meth-analysis}.  

\subsection{Thematic Analysis: Axial Coding and Codebook Development}
\label{meth-analysis}

We transcribed the data from both the comment sections and interview sessions and analyzed them separately. Each dataset was systematically reviewed using \textit{Atlas.ti} to identify recurring themes and patterns. Following Strauss and Corbin's approach to axial coding \cite{strauss1990basics}, we grouped the data codes into overarching themes. Comment and interview data were independently organized into activities, and each activity, corresponding to a code, was assigned to a sensemaking activity pattern, corresponding to a theme. These themes were then assigned to key themes, for which we implemented established sensemaking activity clusters \cite{koesten_talking_2021}: \textit{inspecting}, \textit{engaging} content and \textit{placing}. We introduced an additional cluster, \textit{responding personally}, to encompass affective activities. The terms within this cluster are grounded in \textit{Bloom's Affective Taxonomy}, where visualization viewer activities are categorized as \textit{perceiving}, \textit{responding}, \textit{valuing}, and \textit{believing}.

\begin{figure*}[t]
    \centering
    \includegraphics[width=435pt]{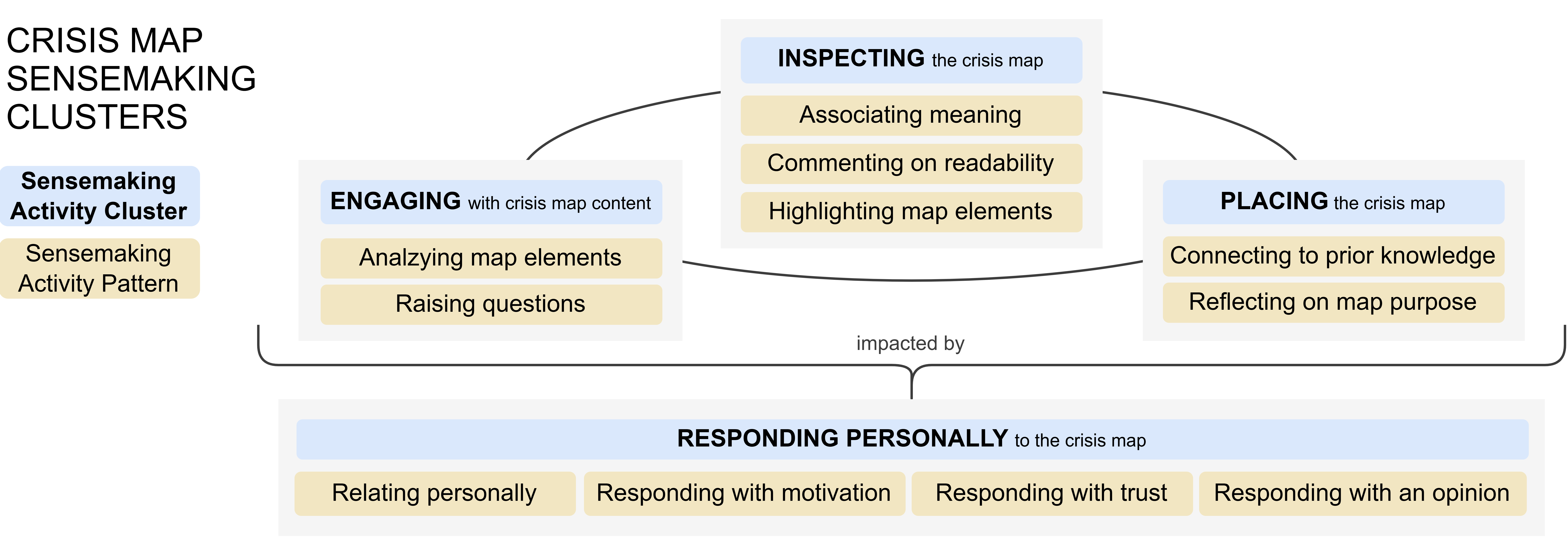}
    \caption{Overview on crisis map sensemaking broken down into sensemaking activity clusters. Each cluster consists of activity patterns which were derived from the thematic analysis of our comment and interview data. A zoomed-in version, where the sensemaking activities are listed for each pattern, follows below (see \autoref{fig:overview-zoomed-in}).}
    \Description[Overview on crisis map sensemaking broken down into sensemaking activity clusters. Each cluster consists of activity patterns which were derived from the thematic analysis of our comment and interview data.]{Overview on crisis map sensemaking broken down into sensemaking activity clusters. There is a text box for each cluster, which consists of activity patterns also portrayed as text boxes. The activity patterns were derived from the thematic analysis of our comment and interview data. The cluster 'inspecting' includes the patterns 'associating meaning', 'commenting on readability', and 'highlighting map elements'. The cluster 'engaging' includes 'analyzing map elements' and 'raising questions'. The cluster 'placing' includes 'connecting to prior knowledge' and 'reflecting on map purpose'. The fourth cluster, 'responding personally', includes the patterns 'relating personally', 'responding with an opinion', 'responding with motivation', and 'responding with trust'.}
    \label{fig:overview-zoomed-out}
\end{figure*}

The axial coding process involved three rounds: an initial round for code emergence, a second round to align phrasing of the codes across studies where semantically appropriate, and a third round aimed at targeted synthesis to categorize the codes as either frictional or non-frictional. Therefore we could assign frictional activities to overarching themes that we call \textit{friction points}, which are inherent to the sensemaking process and occur in connection to sensemaking activity clusters. The systematic data review, coding, and theme assignment were conducted by two of the authors and refined in regular discussions with two senior authors. This process was carried out independently for each study, with only the phrasing of codes aligned in the second round of coding, and the resulting codebooks are provided in the supplementary material (see Figure S1 and Figure S2 for the comment analysis and see Figure S3 to Figure S7 for the interview analysis). The codebooks detail the sensemaking activities, their assignment to activity clusters, and the identification of friction points.

\section{Findings}
We present an overview of crisis map sensemaking (\autoref{fig:overview-zoomed-out}), integrating the emerging sensemaking activity patterns identified across both studies. In \autoref{fig:overview-zoomed-out} sensemaking is broken down into activity clusters with specific activity patterns, and it introduces \textit{responding personally} as an additional activity cluster, where viewers' actions are distinctly affective and, for example, influenced by motivation, trust, or prior beliefs. As noted in \cite{koesten_talking_2021}, the clusters \textit{inspecting}, \textit{engaging}, and \textit{placing} are interconnected and transition fluidly. While we did not focus on the sequence of activities in crisis map sensemaking, our analysis revealed that the \textit{responding personally} cluster often intertwined with these established clusters, such as expressing concern during \textit{inspecting} or sharing experiences during \textit{placing}.

In the following, sensemaking activities are described separately for each study, due to the studies' different contexts and methodologies. Following the description of the sensemaking process in each study, we go on to highlight specific friction points that occurred as part of crisis map sensemaking. These points synthesize findings on frictional sensemaking activities from both studies.

\subsection{Crisis Map Sensemaking broken down into Activity Clusters}
\label{results-activity-clusters}
We analyzed how young, digitally native viewers make sense of crisis maps by clustering their activities into: \textit{inspecting}, \textit{engaging with content}, \textit{placing}, and \textit{responding personally}. Below, without directly comparing the studies, we describe each sensemaking activity cluster, outline subordinate patterns with specific activities, and provide illustrative quotes. The description of sensemaking activities in both studies also includes those identified as frictional during axial coding. These activities are mentioned here as part of the sensemaking process but will be synthesized in Section \ref{results-friction} and further discussed in Section \ref{disc-friction}.

\subsubsection{In the Thematic Analysis of Online Comments}
\label{results-study1}
As viewers \textbf{inspected the crisis maps}, they showed initial reactions and associations and highlighted map elements. There was a tendency to provide a brief introduction to the map and its topic, and to refer to the map title or color usage. Upon first viewing, the implementation of color was frequently mentioned, such as by outlining it: \textit{"[There is] a lot more red, significant amount of pink, too"} (MAP5)\footnote{Comment authors are anonymous. The number after "MAP" indicates the tested crisis map to which the citation refers. All $13$ tested crisis maps are described and linked in Table S1 in the supplementary material.}. As viewers inspected, they also associated meaning, for example by sharing an opinion or by commenting on the perceived relevancy of the depicted crisis: \textit{"The map [...] acknowledges a problem, one that needs to be addressed and figured out soon"} (MAP12).

\begin{figure*}[b]
    \centering
    \includegraphics[width=1\textwidth, keepaspectratio]{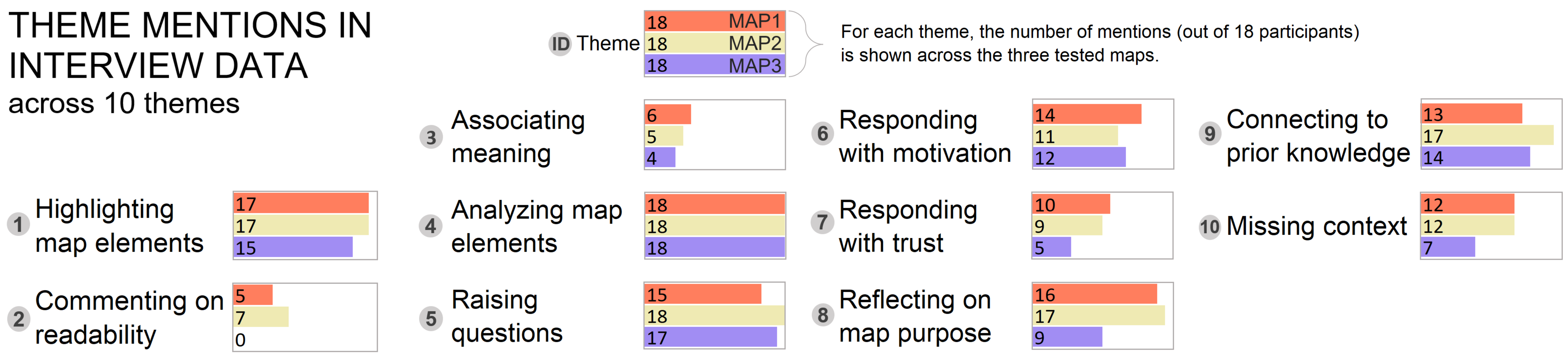}
    \caption{The distribution of theme mentions in the interview study is shown across 10 themes, each represented as a bar chart. Bars indicate the number of mentions (out of 18 participants) for a tested crisis map. Theme descriptions and IDs (1–10) are detailed in Section \ref{results-study1}. A table version is available in Table S3 of the supplementary material.}
    \Description[Figure 3 visualizes the distribution of mentions across ten themes in the interview data, categorized by three tested maps (MAP1, MAP2, MAP3) and for 18 participants. For example, all 18 participants consistently analyzed map elements across all three maps (Theme 4).]{Overview on the distribution of mentions for ten themes identified in the interview data, distributed across three tested maps (MAP1, MAP2, MAP3) and covering the data from 18 participants. The themes include 'Highlighting map elements,' 'Commenting on readability,' 'Analyzing map elements,' 'Associating meaning,' 'Raising questions,' 'Responding with motivation,' 'Responding with trust,' 'Reflecting on map purpose,' 'Connecting to prior knowledge,' and 'Missing context.' The number of mentions for each theme is represented as horizontal bars, color-coded by map. For example, Theme 4 ('Analyzing map elements') shows all 18 participants analyzing map elements consistently across all three maps. The table version of this figure is available in Supplementary Material Table S3.}
    \label{fig:themes-distribution}
\end{figure*}

When viewers \textbf{engaged with the crisis maps}, this encompassed typical steps of map analysis, such as examining value distribution, identifying trends or patterns, comparing variables, and grouping items spatially. Often, they raised questions by wondering about values that stood out to them: \textit{"I wonder why precipitation had a major increase over the last thirty years"} (MAP8), or how the depicted data came to be: \textit{"I wonder how this graph was made. How did the researchers come up with the needed power for 2050, and how did they decide where the power would come from?"} (MAP7). Some viewers made premature assumptions, such as mistaking predictions for facts. For example, MAP7, which was based on models for future wind and solar power needs in the United States, was mistakenly interpreted as a certain fact. Throughout engaging, they frequently focused on color usage in the map. Referencing the implementation of saturation levels or different color hues was used for pointing out areas on the crisis maps: \textit{"[the southwestern] area isn't very blue"} (MAP8).

As viewers \textbf{placed the crisis maps}, they delved deeper into the impacts and effects of depicted crises, sometimes while deriving causes or discussing future scenarios. Generally, they contemplated the map's message, such as for MAP5, which showed endangered biodiversity across the United States: \textit{"this reflects on how our local policies impact our biodiversity and how we protect species that are endangered"}, or for MAP10 on air pollution deaths in the United States: \textit{"graphs like these are absolutely vital to maintaining good health"} (MAP10).  Viewers connected the map to their prior knowledge, such as on demographics or economic structures. Occasionally, viewers reconsidered assumed connections when their knowledge did not apply appropriately. While placing, viewers \textbf{responded personally} by questioning their personal crisis responsibility or assessing their own level of risk. Further, they related the data or the map's message to their place of residency: \textit{"we can learn more about our community's air quality"} (MAP10). As they related, viewers also shared personal experiences, such as for MAP6, which showed extreme temperatures in the United States: \textit{"a lot of things were damaged from the heat and I was worried about people who don't have the advantage of an AC [...] to keep them cooled off"} (MAP6). Some viewers responded with motivation: \textit{"I want to help!"} (MAP12), or with emotion by sharing empathetic thoughts: \textit{"I wonder how the countries that are in the red are feeling right now, I could never imagine what is going through their minds"} (MAP3). If challenges arose as viewers placed a map, this led to rethinking interpretation and re-engaging or re-inspecting. This dynamic appears once again in the follow-up interview study to the comment analysis, shown in Section \ref{results-study2}. 

\subsubsection{In the Interview Study}
\label{results-study2}

As viewers \textbf{inspected the crisis maps}, they highlighted map elements \includegraphics[height=\fontcharht\font`\B]{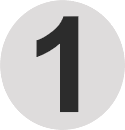} (Circled numbers indicate a cross-reference to theme IDs in \autoref{fig:themes-distribution}.), such as the title and map legend. Similar to the comments from the prior study, there was a tendency to introduce the map by outlining its topic, while also referring to map elements and color usage: \textit{"So, there is a map of the U.S. with COVID-19 cases in the districts. And they are shown by these red circles"} (P16-MAP1). They commented on readability  \includegraphics[height=\fontcharht\font`\B]{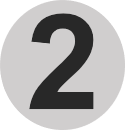}, with some perceiving the maps as cluttered and others as clear. Color usage was frequently mentioned, and sometimes found fitting, such as when contrast was high due to saturated colors, and other times it was found irritating: \textit{"First of all, I feel like it's a lot, and it looks so messy at first with all the red circles"} (P10-MAP1). Viewers shared their overall impression of the crisis maps upon initial viewing, sometimes while associating meaning {\includegraphics[height=\fontcharht\font`\B]{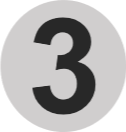}. In the case of MAP1, which showed the number of COVID-19 cases by U.S. county in 2020, three viewers immediately perceived the map as a warning, among other things due to the usage of signal colors. For MAP2, which showed global water stress levels in urban areas, two participants immediately associated it with the topic of climate change: \textit{"So at the very beginning, when [...] the graphic [was first shown], climate change popped into my head"} (P9-MAP2), as one of them (P9) mentioned, this was due to the currentness and urgency of this issue.

Consistently, viewers \textbf{engaged with crisis maps} by analyzing map elements \includegraphics[height=\fontcharht\font`\B]{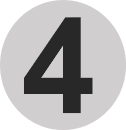}, noting any lack of understanding, and raising questions \includegraphics[height=\fontcharht\font`\B]{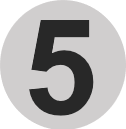}: \textit{"There is something missing for me to have a logical connection. There's a gap in my mind. And no matter how much it's whirring right now, I can't figure out what really...what does the title really mean?"} (P7-MAP3). Viewers struggled with specific map elements such as overlapping symbols, absolute values, insufficient labels for the color key, lack of variable explanation and a "missing data fields" section. In MAP2 on global water stress levels, 13 out of 18 participants found the missing data fields ambiguous: \textit{"I don't know what's behind the gray fields. Either everything is perfect, or it's going so well that they don't need any water, [...] I can only speculate"} (P7-MAP2), and 8 out of 18 participants lacked a definition of water stress as a variable: \textit{"Is it water stress of drinking water, water stress for agriculture, water stress in general for the economy, or something else? Does water stress consider highly seasonal or time-limited dry periods? [This] cannot be inferred from the map at all"} (P16-MAP2). 

Some viewers found it easier to understand elements they perceived as familiar, like growing circles in MAP 1. Most viewers explored the spatial distribution of values, and some prioritized elements: \textit{"Now I notice that I pay less attention to the size of the circles than to the intensity of the color"} (P2-MAP1). Throughout engaging, viewers posed questions and sought answers, sometimes by making assumptions. Viewers focused on familiar regions, and some \textbf{responded personally} with affection towards impacted areas on the map. Viewers mentioned motivational factors, such as personal interest in a crisis issue, that influenced their level of engagement \includegraphics[height=\fontcharht\font`\B]{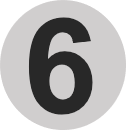}: \textit{"The population development of the USA is not something that personally interests me, so I wouldn't further engage with it"} (P18-MAP3). Some participants were motivated by the map's visual appeal, while others were engaged because they felt that the displayed issue was current and relevant. Some viewers found the crisis issue overly covered in the media and were therefore disinterested. Trust also played a role during sensemaking \includegraphics[height=\fontcharht\font`\B]{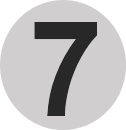}, and was influenced by factors like data accuracy and reliability. 

Viewers \textbf{placed the crisis map} by reflecting on its purpose \includegraphics[height=\fontcharht\font`\B]{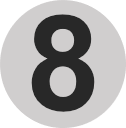}. They contemplated messages, implementation scenarios, and evaluated the design. The interpretation of the map's purpose varied, but it was commonly seen as a geographical overview and a means to raise awareness about the crisis: \textit{"The color selection [is][...] a very strong signal [...]. To describe the seriousness of the situation, [the map] was definitely a good representation"} (P15-MAP1). Some saw potential for the maps as decision-making tools, useful to government officials. Others \textbf{responded personally} to the crisis maps, for example by feeling personally addressed to save water: \textit{"We have a water shortage that already exists and is getting worse, so we should simply be mindful not to waste water"} (P1-MAP2). Viewers used their prior knowledge \includegraphics[height=\fontcharht\font`\B]{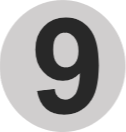} to contextualize or verify their interpretations. Sometimes they confirmed assumptions, other times they were surprised but adopted new perspectives: \textit{"I'm a bit surprised that there are certain areas that are in the lower range. I didn't imagine it that way, but yes"} (P4-MAP2). Some struggled with unfamiliar issues and noted the risk of misinterpretation without adequate prior knowledge: \textit{"I believe the map can also be easily misunderstood, for example, if the title is misinterpreted"} (P2-MAP1). Often, viewers mentioned that there was missing context for the crisis maps \includegraphics[height=\fontcharht\font`\B]{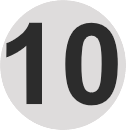} and felt therefore limited in drawing robust conclusions from the crisis maps. Missing context was a key friction point, alongside other issues, which are detailed in Section \ref{results-friction}.

\subsection{Synthesizing the Studies: Friction Points in Crisis Map Sensemaking}
\label{results-friction}
We identified four friction points as inherent parts of crisis map sensemaking, drawing on the results from both studies: struggling with color encoding, missing context, lacking connection, and distrusting the map. These friction points emerged through a synthesis of the frictional activities in the codebook data. Our understanding of these friction points is informed by the theory of Threshold Concepts' liminal space, where learners may experience difficulties or get stuck before acquiring new knowledge \cite{schwartmann2010thresholdconcepts, goebel2019thresholdconcepts}. A detailed overview (\autoref{fig:overview-zoomed-in}) of sensemaking activities, their assignment to clusters, and friction points is provided and further discussed in Section \ref{friction-sensemaking}.

\subsubsection{Struggling with color encoding} Various design choices for map elements in the surveyed crisis maps influenced efficient sensemaking, with color encoding being the most challenging. There were difficulties understanding aspects of color as an encoding channel, especially when viewers' semantic associations of color did not match the represented information: \textit{"I thought [the colors] had something to do with heat because red and yellow-red are colors associated with heat. [...] [N]ow I read that it's about water stress, so my assumption wasn't correct at all"} (P13-MAP2). However, issues with comprehending color encodings were not always uncovered but sometimes led to inaccurate conclusions. Such as mistaking negative growth for a positive trend due to misread divergent color usage, like in MAP8, where U.S. precipitation was shown over time, and the decrease in precipitation (marked in light yellow to muddy brown) was often confused with meaning drought. Sometimes, viewers criticized a lack of information, even though it was presented on the map but not recognized. They just did not decode its representation through color. Viewers found guidance insufficient, and many suggested more detailed descriptions. Consequently, some viewers proposed changes to the color design, suggesting colors that better align with themes or evoke specific associations based on prior experiences.

\subsubsection{Missing context} Viewers often mentioned a lack of context in the crisis maps and, therefore, perceived them as difficult to read: \textit{"having some text or information beforehand, or telling people what it's about, would be helpful. [...] [R]ight now, I find it a bit difficult to understand”} (P6-MAP3). Viewers were missing details like a publication date, the data collection period, or publication context. The absence of this information complicated sensemaking and led viewers to rely on prior knowledge or make assumptions. They desired additional information, especially in textual form, to enhance comprehensibility: \textit{"it might be different if [the map] were explained in a text. Like, why or how the developments happened, [...] a short text to get familiar with it, so that I really understand it"} (P3-MAP3).

\subsubsection{Lacking connection} Viewers felt disconnected from the crisis maps for several reasons. A lack of expertise in the depicted topic or geographic information led to uncertainty and self-doubt regarding map reading ability. Geographical knowledge was an issue, with difficulty identifying countries and regions, partly due to the absence of orientation aids like city names and country labels: \textit{"It’s not clear to me if 'county' is the same as 'state'. Probably not. Maybe 'county' refers to each small village. Oh, I don’t know. This should be clear to me, definitely. Maybe I’m just uneducated about this"} (P17-MAP1). Some viewers stressed their disconnection from the displayed crisis issue: \textit{"This really doesn't affect my community because there's rarely any change in California"} (MAP7). Viewers were also challenged by demotivation when they perceived maps as difficult, when the crisis issue did not feel current, or when they found the crisis issue over-communicated.

\subsubsection{Distrusting the map} Viewers faced challenges with trusting the crisis maps due to perceived issues with data transparency, detail, and sourcing. They desired more information on data collection, processing, and publication context. Distrust was also triggered by the omission of certain areas on the maps: \textit{"My trust diminishes because something is missing. Actually, the lack of information raises suspicion"} (P7-MAP2). While some viewers distrusted the maps for the aforementioned reasons, others found them trustworthy due to their sourcing, or the transparent handling of missing data: \textit{"Certain information or data missing could be an indication that the data depicted on the map are accurate. It is possible that someone creating a map would not use unreliable data or invent data and insert them"} (P2-MAP2). In general, viewer trust was influenced by personal responses, as for example the handling of missing data was judged from a personal point of view.

\begin{figure*}[ht]
    \centering
    \includegraphics[width=430pt]{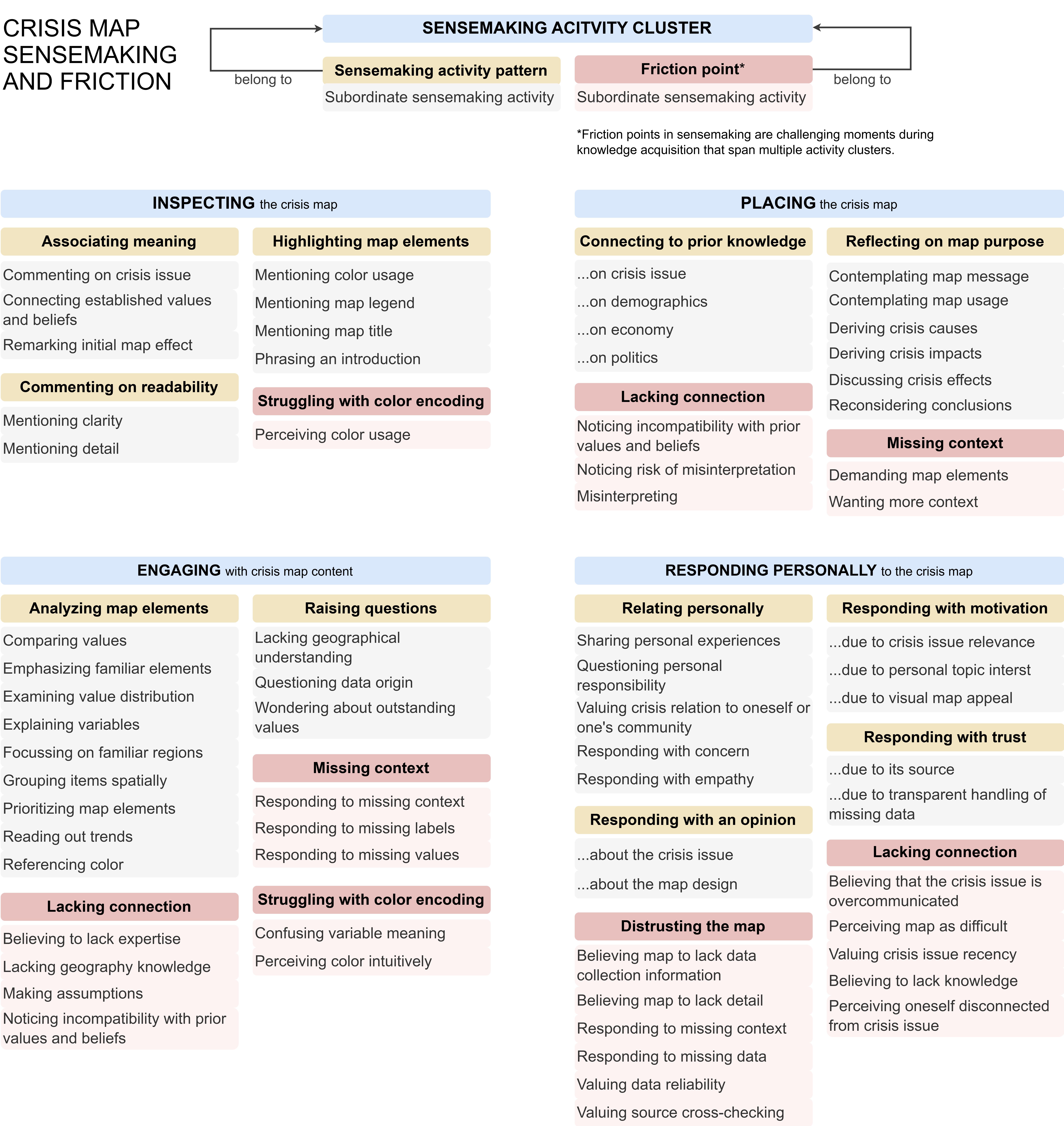}
    \caption{Overview of crisis map sensemaking, including friction points and detailed sensemaking activities. Activities derived from both studies are grouped into overarching patterns or friction points, which are connected to broader activity clusters. A table version is available as Table S4 in the supplementary material.}
    \label{fig:overview-zoomed-in}
    \Description[Crisis map sensemaking overview including friction points and zoomed in to show sensemaking activity level. Sensemaking activities are based on the analysis of comment and interview data. They were assigned to overarching sensemaking activity patterns or friction points, both of which are connected to overarching activity clusters.]{Figure 4 shows a crisis map sensemaking overview, including friction points and sensemaking activities. Sensemaking activities are based on the analysis of comment and interview data. This is a zoomed-in view of Figure 3. Activities were assigned to overarching sensemaking activity patterns or friction points, both of which are connected to overarching activity clusters. All elements are shown in text boxes. For the pattern 'associating meaning', for example, the activities commenting on crisis issues, connecting to established beliefs, and remarking on the initial map effect are listed. In total, 69 activities are shown in the overview. Friction regarding struggles with color encoding occurs in relation to the clusters 'inspecting' and 'engaging'. Friction regarding a lack of connection occurs during 'engaging', 'placing', and 'responding personally'. Friction due to missing context occurs in 'engaging' and 'placing'. Friction due to distrust is connected to 'responding personally'. The table version of this figure is available in Supplementary Material Table S4.}
\end{figure*}

\section{Discussion}
In this paper, we drew on a mixed methods study, incorporating a comment analysis and semi-structured interviews, to explore crisis map sensemaking by Digital Natives. To address our aim of understanding what challenges tell us about the sensemaking process, the discussion is framed through the lens of friction. We identified four key friction points when understanding crisis maps—struggles with color encoding, missing context, lack of connection, and distrust—each part of different stages of the sensemaking process. Here, we position these points within the sensemaking process \autoref{fig:overview-zoomed-out} and connect them to sensemaking clusters identified in \autoref{fig:overview-zoomed-out}. We discuss each point in relation to relevant literature and outline implications based on our findings.

\subsection{Placing Friction Points in the Data-Centric Sensemaking Framework}
\label{friction-sensemaking}

Friction points in sensemaking describe "troublesome" moments during knowledge acquisition, as defined in the theory of Threshold Concepts \cite{schwartmann2010thresholdconcepts, goebel2019thresholdconcepts}. These points, rather than being negative, may accelerate or deepen understanding \cite{koesten_talking_2021, boukhanovsky2017uncertainty}, and are, therefore, key aspects of sensemaking. In our analysis of crisis map sensemaking, we identified friction points, which consist of specific activities and span multiple sensemaking activity clusters. In \autoref{fig:overview-zoomed-in} we provide an overview of crisis map sensemaking, including the placement of friction points within the process. These friction points are shown alongside sensemaking activity patterns, described in Section \ref{results-study1} and Section \ref{results-study2}. Like the sensemaking activity patterns, which can be broken down into sensemaking activities \cite{koesten_talking_2021}, friction points also consist of specific subordinate activities. For example, the pattern \textit{associating meaning} includes activities like "commenting on the crisis issue" and "remarking the initial map effect" (see \autoref{fig:overview-zoomed-in} for more examples). Similarly, the friction point related to a lack of connection involves "believing to lack expertise" or "perceiving oneself disconnected from the crisis issue" (see \autoref{fig:overview-zoomed-in}). 

In \autoref{fig:overview-zoomed-out} we depicted crisis map sensemaking as a process, based on our findings, showing sensemaking to consist of four activity clusters: \textit{inspecting}, \textit{engaging}, \textit{placing} and \textit{responding personally}. Each cluster consists of distinct activity patterns, such as "analyzing map elements" which is tied to \textit{engaging} (see \autoref{fig:overview-zoomed-out}), or "reflecting on map purpose" which belongs to \textit{placing} (see \autoref{fig:overview-zoomed-in}). As established by \cite{koesten_talking_2021}, sensemaking activity patterns belong to specific clusters, but our results show that friction points do not. Unlike sensemaking activity patterns, friction points span across multiple clusters, rather than being confined to a single one: 

\begin{itemize}
    \item \textbf{Struggling with color encoding} was common during \textit{inspecting} and \textit{engaging}, particularly when viewers highlighted or analyzed map elements, as color often influenced their interpretation. 
    
    \item \textbf{Missing context} caused irritation during \textit{engaging} and \textit{placing}, especially when viewers lacked labels or data, prompting them to raise questions and seek further information.
    
    \item \textbf{Distrusting} the map emerged as a \textit{personal response} to perceived gaps or omissions, particularly when viewers questioned the transparency or completeness of the data presented. 
    
    \item \textbf{Lacking connection} affected \textit{engagement}, \textit{placing}, and \textit{personal response}, with viewers feeling uncertain due to perceived gaps in expertise or knowledge, influencing their ability to relate to and reflect on the map’s purpose.
\end{itemize}

\subsection{Sensemaking Friction Points and Their Implications}
\label{disc-friction}

Addressing friction points can help map readers to more effectively navigate the liminal space between confusion and understanding, and is therefore essential for enhancing the perceived reliability and credibility of crisis maps. Our findings raise awareness of how crisis maps are made sense of, particularly by a young audience. They further have concrete implications for the design of effective modes of communicating crisis information visually. Below, we connect identified friction points to existing literature and address key actors involved in the design and use of crisis maps.

\subsubsection{Considering color associations} Color usage has been shown to be significant for visualization perception, as viewers are influenced by brightness, saturation, and hue choices \cite{munzner_visualization_2014}. In both the comment analysis and the interviews, viewers often referenced regions by colors and highlighted their semantic understanding of color hues, which have been shown to be crucial for first visualization impressions \cite{hogan_elicitation_2016, fan_understanding_2022} and for cartography in general \cite{bertin_semiology_1983, zhang_mapping_2021, fang_evaluating_2022, cay_understanding_2020}. Warm tones may enhance visual prominence and speed of information assimilation \cite{fang_evaluating_2022}, such as red, which is commonly used as a warning color and is likely to be familiar to viewers \cite{kaufmann_communicating_2020, griffith1997association}. However, our findings indicate that such commonly used colors might not always be optimal. In the case of MAP2, the usage of red led to confusion in the interviews, as it represented water stress, but was often mistaken for representing heat or drought. Similar associations exist for other colors, such as for blue increasing trust in viewers \cite{su2019blue}. Hence, it is important to consider potential mismatches between color associations and variables \cite{szafir_modeling_2018}.\\

\textbf{Implications}: Map designers, news organizations, and researchers should test color associations with target audiences during the design process to identify potential misunderstandings or usability issues. This is particularly important in crisis maps where misinterpretation may prevent crucial information retrieval \cite{buckingham2009future, kovalchuk2023digitalization}, as seen in MAP2, where red was mistaken for heat when it represented water stress. Testing color choices through focus groups or pilot studies can uncover these mismatches early. An iterative feedback loop, like usability testing, could help designers refine color schemes based on user input. Moreover, this process could include an assessment of semantic associations of color, considering how it may also carry implicit meanings for different audiences. For example, in MAP8, divergent color scales were misread as representing drought conditions, stressing the need for clearer cues and context-specific guidance.

\subsubsection{Providing sufficient context} We found that viewers were often missing context, which has been shown to potentially lead to significant misinterpretations \cite{shklovski_finding_2008, hiltz_dealing_2013, oh_community_2013}. For instance, an analysis of \textit{Twitter}\footnote{Formerly \textit{Twitter}, now called \textit{X}.} comments on crisis maps shows that accurate visualizations can inadvertently support misinformation when presented without proper context \cite{lisnic_misleading_2022}. In our interview study, viewers suggested complementing maps with other forms of information transmission. This aligns with research indicating that various forms of representation, such as text, diagrams, or interactive visualizations, enhance comprehensibility and information absorption \cite{hullman_benefitting_2011}. Other work supports the need for a balance between text and visual elements \cite{cleveland1984graphical, kosslyn1989understanding, pinker1990theory}, which aligns with viewer demands in our interview study for more elaborate text elements.\\

\textbf{Implications}: News providers often enhance maps with annotations or explanatory texts, which can support viewer interpretation \cite{kalir2021annotation}. Content moderators might consider overseeing critical issues on social media to ensure that viewers perceive and integrate annotations. At the same time, content moderators may step in to provide clarifications and address misunderstandings directly in the comment sections \cite{dailey2015collectivesensemaking}, as to prevent spread of misinformation \cite{orso2020infodemic, lisnic_misleading_2022}. Educators can also help by teaching students to critically analyze visualizations. For instance, comparing maps with and without detailed annotations can illustrate how supplementary information affects understanding. Friction caused by missing context in crisis maps parallels sensemaking challenges in other areas, such as in interpreting search results \cite{yihan2013sensemakinginfo}.

\subsubsection{Building suitable viewer connections} In our studies, viewers related the crisis maps to themselves. They connected their personal experiences, assessed risks in areas related to their lives, expressed opinions, and connected personal associations. This aligns with studies showing that a viewer's proactive search for personal associations in maps is used to verify information and reinforce their understanding of a visualization \cite{hogan_elicitation_2016, nowak_micro-phenomenological_2018}. Research in proximity techniques has shown that viewer interest is enhanced through perceived relevancy of visualized data \cite{campbell2019feeling}, and that the viewer's feeling of data, such as understanding how a crisis might affect their local community or themselves, influences their sensemaking \cite{kennedy2018feeling}. In both of our studies, we found that a lack of personal relevance reduced viewer engagement, aligning with findings that proximity boosts engagement with crisis topics online \cite{zhou_spotlight_2023}. However, our findings also show that when participants strongly identified with the topic, their focus on personal connections sometimes distracted them from the displayed data. This aligns with previous research emphasizing that a 'one-size-fits-all' approach is ineffective for diverse users with varying map literacy and contextual knowledge \cite{kostelnick_cartographic_2013, skinner_physicians_1994}.\\

\textbf{Implications}: Map designers should aim to create a balance, targeting audience interest without overwhelming or alienating them. Maps that are visually complex, for example with a large number of data layers or dense information, can cause cognitive overload \cite{harold2019} -- on the other hand, overly simplistic maps might fail to communicate the nuances of a crisis situation. Crisis maps that allow users to personalize their experience could enhance relevance and engagement, especially for digitally native audiences. Though this audience group is not monolithic \cite{correa2016digital}, they tend to be familiar with digital environments \cite{underwood2007rethinking}. Our study found that viewers focused on familiar areas but were frustrated by missing information about unfamiliar areas, suggesting that localized information based on a viewer's location, community, or personal interest could boost engagement.

\subsubsection{Designing transparent and trustworthy maps} The influence of visualizations, including maps, on trust and decision-making is well-documented \cite{griffin_trustworthy_2020, monmonier_cartographies_1997, juergens_trustworthy_2020, mayr_trust_2019}. Missing data significantly affects trust in visualizations \cite{gleicher2013perception, pipino2002data, pouchard2015revisiting}, especially when conveying risk \cite{marsh2003role, renn1991credibility}, as also shown in our studies, where viewers judged the trustworthiness based on their stance towards the flagging of missing data. It has been suggested, that informing viewers about data uncertainty increases trust \cite{sacha2015role}, aligning with our findings where participants saw labeled missing data as a sign of honesty and transparency. However, excessive transparency may negatively impact trust \cite{xiong_examining_2019, hood2006transparency}, which was also the case for a share of the viewers. This indicates a need to balance this tension in conveying data uncertainty on maps. Further, we found that lack of context or vague sourcing, such as missing data labels, ambiguous map legends, or unclear geographical markers, also causes distrust.\\

\textbf{Implications}: Crisis communicators should balance transparency with clarity, avoiding overwhelming viewers while providing sufficient context to aid understanding. For example, maps that include too much technical jargon can confuse audiences \cite{xexakis2021}, who might lack the expertise to interpret such details. To avoid this, interface designers could enhance user trust by incorporating features that allow viewers to easily verify and evaluate data shown on the map. Interactive layers, where users choose different levels of detail, could help ensure that critical information is accessible while providing further insight for those who seek more technical data. Additionally, details on data provenance, collection methods, and uncertainty should be provided on demand, as their absence irritated viewers or even caused distrust. For instance, participants expressed frustration when no data origin was indicated for predicted data in MAP7. Distrust might be more pronounced in crisis maps than in other thematic maps due to heightened stakes and the need for viewers to rely on data for critical decision-making. The urgency of crisis-related topics might drive how viewers approach crisis maps and bring specific expectations of accuracy that differ from other thematic visualizations.

\subsection{Future Work}
Future research on crisis maps could address the identified sensemaking friction points by exploring specific aspects such as map context or strategies for building viewer connection. For example, the use of data overlays with background information, or showing proximity on crisis maps can be tested with different levels of embodiment. In future endeavors, interactive visualization which are part of current information culture \cite{jia2024interactive}, could provide additional insights into activities relevant to sensemaking, for example by investigating the process when users engage with features like adjustable layers or guided narratives. Comparisons between static crisis maps and interactive systems may reveal unique sensemaking activities and friction points or validate the applicability of our findings across domains. Further, while some findings may apply to other thematic maps, their manifestations and impact might differ. Testing in broader visualization contexts could help identify which sensemaking activities and friction are generalizable and which are crisis-specific.

Our findings could be expanded by examining sensemaking on online platforms where crises are communicated, such as government and public health websites, online encyclopedias, and educational resources. Real-time visualizations, such as live public health dashboards, present another area of interest, particularly in exploring how dynamic contexts affect friction points like distrust. Additionally, research could explore the sensemaking processes of other groups beyond young, digitally native viewers, such as working professionals with non-technical backgrounds or K-12 students with lower visual literacy. Collaborative sensemaking, for instance in examining sensemaking activities in digital spaces like crisis communication forums, is another promising direction. Future studies might also focus on specific crisis issues rather than general crisis-related maps, allowing researchers to gain more nuanced insights into how particular crisis topics affect sensemaking and to differentiate personal responses to maps. 

\section{Limitations}
\label{limitations}

In our studies, comments and interview results might be influenced by social desirability biases. There are also limitations to exploring real-time map sensemaking. Though the employment of tasks during sensemaking is a common approach \cite{hogan_elicitation_2016, cay_understanding_2020, nowak_micro-phenomenological_2018}, they may have influenced participant behavior. In the first study, comments were shaped by the four-question structure of the \textit{New York Times} series, often leading to repetitive phrasing, like “I wonder...”. While setting a minimum word count ($100$ words) for comment selection ensured substantivity, it introduced a selection bias by potentially excluding shorter but meaningful responses. This criterion may have limited the dataset to those more comfortable expressing themselves in writing. Future analysis could address this by including shorter comments, guided by supplementary criteria like thematic relevance. Additionally, direct authorship and location of the commenters could not be verified, as they submitted comments individually or in classroom settings, and the provision of personal details was not consistent across all comments. In the second study, interviews might have been influenced by participants' comfort with the think-aloud method.

Sample biases could also arise from the distinct geographical, cultural, and socioeconomic backgrounds of the participant groups. The responses in the comment section reflect a specific U.S.-centric, high school context, which could limit generalizability. The interview study provided a contrasting perspective to the comment sample, by including 18 participants who had varying levels of professional and educational experience (see \autoref{tab:interview-participants}) and lived in three Western European countries. Although we recognize that our research on sensemaking and its friction could be enriched by considering interactive visualizations, we focused on static ones as this format was predominant in the data visualizations provided by the \textit{NYT} (with the exception of MAP12). Furthermore, while our studies relate crisis map sensemaking to correct map understanding, this was not directly measured. Unlike quantitative literacy assessments related to data visualizations like VLAT \cite{kim2017} or CALVI \cite{ge2023}, which use right-or-wrong questions to evaluate data comprehension, we focused on the process of how participants make sense of crisis maps without quantifying their understanding.

\section{Conclusion}
The paper examines how viewers understand crisis maps and the frictions encountered in the process. Through two qualitative studies, we explored the perspectives of young, digitally native viewers, whose comprehension of maps is essential in an era of online information. Our thematic analysis identified four sensemaking activity clusters: \textit{inspecting} the crisis map, \textit{engaging} with its content, \textit{placing} the map, and \textit{responding personally} to the map. The inclusion of personal and affective responses was shown to be a critical part of the sensemaking process. We identified and discussed friction points that have implications for the design and implementation of crisis maps, particularly regarding color encoding, context provision, and fostering viewer connection and trust. Our findings underscore the importance of viewer-centered map designs to ensure that crisis maps effectively communicate critical issues. Additionally, they highlight the need to investigate map perception as a process involving various interacting factors, which should be studied across different audience groups to ensure that crisis maps effectively communicate critical issues.




\bibliographystyle{ACM-Reference-Format}

\bibliography{bib}



\end{document}